\documentclass{article}

\usepackage{arxiv}

\usepackage[utf8]{inputenc} % allow utf-8 input
\usepackage[T1]{fontenc}    % use 8-bit T1 fonts
\usepackage{hyperref}       % hyperlinks
\usepackage{url}            % simple URL typesetting
\usepackage{booktabs}       % professional-quality tables
\usepackage{amsfonts}       % blackboard math symbols
\usepackage{nicefrac}       % compact symbols for 1/2, etc.
\usepackage{microtype}      % microtypography
\usepackage{graphicx}
\usepackage{natbib}
\usepackage{doi}
\usepackage{amsmath}
\usepackage{mathtools}
\usepackage{lscape}
\usepackage{adjustbox}
\usepackage{float}
\DeclareMathOperator*{\argmax}{arg\,max}
\DeclareMathOperator*{\argmin}{arg\,min}
\DeclareUnicodeCharacter{2061}{}

\title{Towards Filling the Gaps around Recurrent Events in High-Dimensional Framework: Literature Review and Early Comparison}

\author{ \href{https://orcid.org/0000-0002-7017-9865}{\includegraphics[scale=0.06]{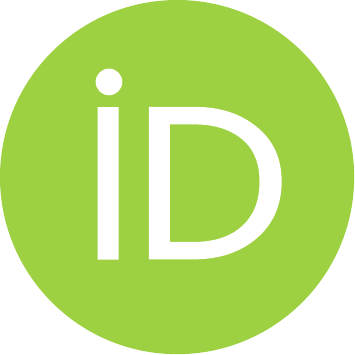}\hspace{1mm}Juliette Murris} \\
	Inserm, Centre de recherche des Cordeliers, Université de Paris, Sorbonne Université, Paris, France\\
	HeKA, Inria, Paris, France \\
	RWE \& Data, Pierre Fabre, Boulogne-Billancourt, France \\
	\texttt{juliette.murris@pierre-fabre.com} \\
	\And
	\href{https://orcid.org/0000-0001-6437-7059}{\includegraphics[scale=0.06]{orcid.pdf}\hspace{1mm}Anaïs Charles-Nelson} \\
	AP-HP, Hôpital Européen Georges-Pompidou, Unité de Recherche Clinique, APHP.Centre, Paris, France \\ 
	Inserm, Centre d'Investigation Clinique 1418 (CIC1418) Épidémiologie Clinique, Paris, France \\
	\And
	\href{https://orcid.org/0000-0002-0049-2397}{\includegraphics[scale=0.06]{orcid.pdf}\hspace{1mm}Audrey Lavenu} \\
	Université de Rennes 1, Faculté de médecine, Rennes, France \\ 
	IRMAR, Institut de Recherche Mathématique de Rennes, Rennes, France  \\
	CIC Inserm CIC 1414, Université de Rennes 1, Rennes, France \\
	\And
	\href{https://orcid.org/0000-0002-7261-0671}{\includegraphics[scale=0.06]{orcid.pdf}\hspace{1mm}Sandrine Katsahian} \\
	Inserm, Centre de recherche des Cordeliers, Université de Paris, Sorbonne Université, Paris, France\\
	HeKA, Inria, Paris, France \\
	AP-HP, Hôpital Européen Georges-Pompidou, Unité de Recherche Clinique, APHP.Centre, Paris, France \\ 
	Inserm, Centre d'Investigation Clinique 1418 (CIC1418) Épidémiologie Clinique, Paris, France \\
	HEGP, Service d'informatique médicale, biostatistiques et santé publique, AP-HP, Paris, France \\
}

\date{}

\renewcommand{\shorttitle}{Murris, \textit{et al.} (2022)}

\hypersetup{
pdftitle={Murris2022},
pdfsubject={q-bio.NC, q-bio.QM},
pdfauthor={Juliette Murris, Anaïs Charles-Nelson, Audrey Lavenu, Sandrine Katsahian},
pdfkeywords={Recurrent Events, Survival Analysis, High-Dimensional Data, Machine Learning},
}

\begin{document}
\maketitle

\begin{abstract}
	\textbf{Background} Study individuals may face repeated events overtime. However, there is no consensus around learning approaches to use in a high-dimensional framework for survival data (when the number of variables exceeds the number of individuals, i.e., $p > n$). This study aimed at identifying learning algorithms for analyzing/predicting recurrent events and at comparing them to standard statistical models in various data simulation settings. \textbf{Methods} A literature review (LR) was conducted to provide state-of-the-art methodology. Data were then simulated including variations of the number of variables, the proportion of active variables, and the number of events. Learning algorithms from the LR were compared to standard methods in such simulation scheme. Evaluation measures were Harrell’s concordance index, Kim’s C-index and error rate for active variables. \textbf{Results} Seven publications were identified, consisting in four methodological studies, one application paper and two reviews. The broken adaptive ridge penalization and the RankDeepSurv deep neural network were used for comparison. On simulated data, the standard models failed when $p > n$. Penalized Andersen-Gill and frailty models outperformed, whereas RandkDeepSurv reported lower performances. \textbf{Conclusion} As no guidelines support a specific approach, this study helps to better understand mechanisms and limits of investigated methods in such context.
\end{abstract}

% keywords can be removed
\keywords{Recurrent Events \and Survival Analysis \and High-Dimensional Data \and Machine Learning}

\section{Introduction}
\begin{figure}[h]
	\centering
    \includegraphics[width=0.8\textwidth]{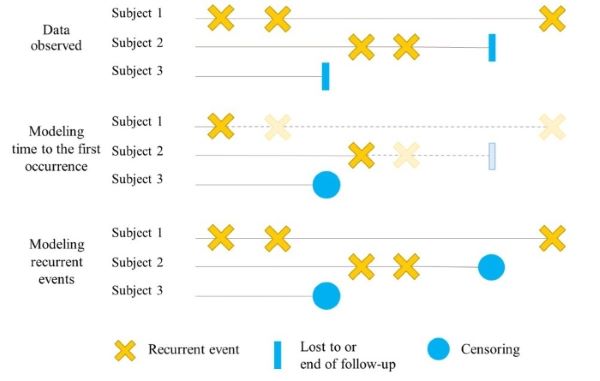}
    \caption{Recurrent Event Framework}
	\label{fig:fig1}
\end{figure}

Study individuals may face repeated events over time, such as hospitalizations or cancer relapses. In either clinical trials or real-world set, survival analysis usually focuses on modeling the time to the first occurrence of the event. Nonetheless, variables may have a varying impact on the first event and on subsequent occurrences. When dealing with multiple occurrences of an event, it becomes a matter of modeling recurrent events (Figure \ref{fig:fig1}). 

Two main challenges arise when analyzing recurrent events. First, interindividual heterogeneity emerges as some subjects may be more likely than others to experience the event. Second, events for an individual are not independent, raising intraindividual heterogeneity. These issues have been statistically addressed through two approaches, marginal and conditional models. Marginal models involve implicitly averaging over the history of previous recurrent events. Conditional models can condition on the history of the events. \\
\\
Moreover, modern technologies enable data to be generated on thousands of variables or observations, as per genomics, medico-administrative databases, disease monitoring by intelligent medical devices, etc. While massive data describe large numbers of observations, high-dimensional data are characterized when the number of variables studied p is greater than the number of individuals n. It is precisely the context of high dimension that will be of particular interest in this paper. Standard statistical models may no longer be applied in this case, as they tend to face convergence problems and non-clinically relevant significance of the variables can arise. To help solve high dimension problems, many machine learning methods have emerged. \\
\\
Literature reviews have previously been conducted on recurrent events, but none has dealt with a high dimensional framework \citep{rogers_analysing_2014, twisk_applied_2005, amorim_modelling_2015}. The objectives of this study were to identify innovative methodology for the analysis and the prediction recurrent events in high-dimensional survival data and to compare them to standard methods using simulated data. \\
\\
The literature review for the identification of innovative approaches is first detailed. Methodology components used for the comparison are then developed and results are outlined. Last sections finally summarize main points of discussion and provide the key messages from this paper.

%%%%%%%%%%%%%%%%%%%%%%%%%%%%%%%%%%%%%%%%%%%%%%%%%%%%%%%%%%%%%%%%%%%%%%%%%%%%%%%%%%%%%%%%

\newpage
\section{Literature review}
\label{sec:LT}

\begin{figure}[h]
	\centering
    \includegraphics[width=0.75\textwidth]{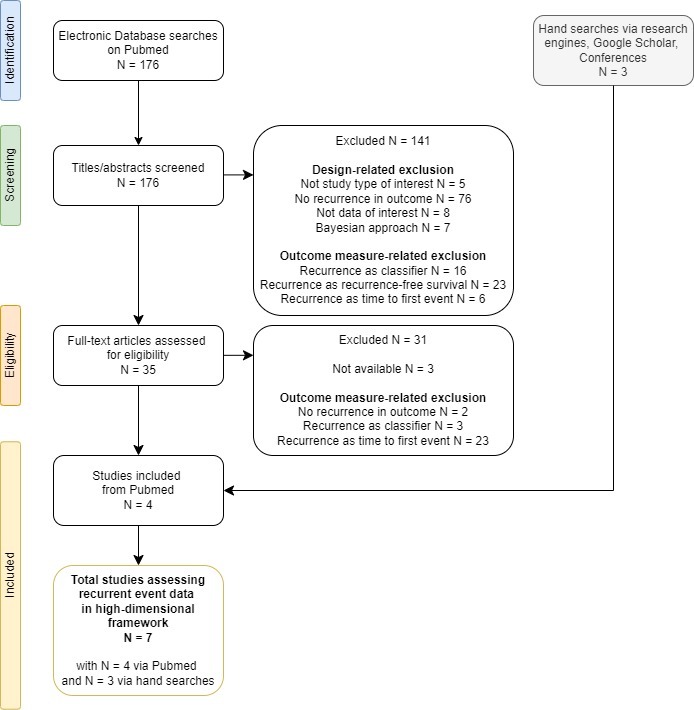}
    \caption{Flowchart of Included Publications via Pubmed}
	\label{fig:fig2}
\end{figure}

A literature review was conducted on PubMed based on Cochrane recommendations \citep{tacconelli_systematic_2010, higgins_cochrane_2019}. Hand searches were then carried out via relevant search engines and conferences \citep{noauthor_guide_2013}. Inclusion criteria were either methodological or observational studies that analyzed any recurrent outcome. Reviews and/or surveys were also eligible. Methods were considered if the publication mentioned that data in a high-dimensional framework were used or if machine learning techniques were employed. Exclusion criteria were any Bayesian approach and clinical trial design. Research equation included (but was not limited to) the following key terms (and associated MeSH Terms): ‘recurrence’, ‘survival analysis’, ‘high-dimension’ and ‘machine learning’. Two reviewers assessed the eligibility of publications independently and any discrepancies were discussed. Data were summarized descriptively based on the following categories: publication and study characteristics, statistical/machine learning approaches used, and application of data. \\
\\
Extraction was performed in November 2021 and led to the identification of 176 hits through electronic research on Pudmed (Figure \ref{fig:fig2}). Overall, after confirming the outcome of interest dealt with recurrence, the primary reason for exclusion was the non-consideration of recurrent events as time-to-event for each occurrence. Recurrence was considered as a classifier (19/176), as a recurrence-free survival outcome (23/176), or as a time-to-first event (29/176). This may be the illustration of authors' caution when dealing with recurrent events in high dimensions, as no published guidelines or recommendations exist to date. In addition, three full-text articles could not be reviewed as they were not available. After title, abstract and full-text thorough review, four publications were included from the electronic database search. Three additional papers from the hand searches were identified. \\
\\

A total of seven relevant publications were selected, consisting in four methodological studies, one application paper and two literature reviews (Table \ref{tab:table1}). Two articles describing learning algorithms for variable selection strategies were identified. First,  focused on accelerating coefficient estimation with a coordinated descent algorithm and penalizing partial likelihood. \citet{zhao_variable_2018} provided an extension of Ridge penalization for estimating and selecting variables simultaneously. \citet{gupta_cresa_2019} and \citet{jing_deep_2019} developed deep neural networks extensions for the analysis of recurrence, respectively. In addition, \citet{kim_deep_2021} proposed an application paper aiming at estimating time between two breast cancer recurrences. However, the methodology used in the latter was an extension of a recurrent neural network which was unfortunately not published in any peer-reviewed journal. Finally, two literature reviews mentioned learning algorithms for recurrent event data analysis, even though none of them provided guidelines \citep{wang_machine_2019, bull_harnessing_2020}. \\
\\
Findings from this literature review demonstrated the current gap in the literature and vast differences in the context and methods of interest. In particular, not all developed models were based on simulated data, such as \citet{jing_deep_2019}. Also, none of the included publications compared their performance one to another. For instance, variable selection approaches were compared to standard statistical model only, while neural networks were compared to other neural networks or random forests. No head-to-head comparison across standard methods, learning algorithms and deep neural network seemed to have been performed. \\

\newpage
\begin{landscape}
    \begin{table}
    	\caption{Papers Identified from the Literature Review}
        \small
    	\centering
        \begin{tabular}{llllllllll}
            \toprule
            \#  & Year & Author & Title & Type & Description 
            & \begin{tabular}[c]{@{}l@{}}Data used for \\application\end{tabular}  & Evaluation measure 
            & \begin{tabular}[c]{@{}l@{}}Code availability \\/ reproducibility\end{tabular}  \\
            \hline \hline
            \#1 & 2013 
            & \begin{tabular}[c]{@{}c@{}}Wu, \\\textit{et al}.\end{tabular} 
            &   \begin{tabular}[c]{@{}l@{}} 
                    Lasso penalized \\semiparametric \\regression on \\high-dimensional \\recurrent event data \\via coordinate descent 
                \end{tabular}
            &   \begin{tabular}[c]{@{}l@{}}
                    Variable \\ selection
                \end{tabular} 
            &   \begin{tabular}[c]{@{}l@{}} 
                    Regularization method with \\penalization and use of the \\coordinated descent algorithm, \\which computes  the deviations of \\the optimization problem and \\updates the parameter value \\iteratively
                \end{tabular} 
            & \begin{tabular}[c]{@{}l@{}}Chronic septic \\granulomatosis \end{tabular} 
            & \begin{tabular}[c]{@{}l@{}}
                    Number of selected \\predictor variables and \\regression coefficients
                \end{tabular} 
            & No \\
            \hline 
            \#2 & 2018 
            & \begin{tabular}[c]{@{}c@{}}Zhao, \\\textit{et al}.\end{tabular}
            &   \begin{tabular}[c]{@{}l@{}} 
                    Variable selection for \\recurrent event data with \\broken adaptive ridge \\regression 
                \end{tabular} 
            &   \begin{tabular}[c]{@{}l@{}}
                    Variable \\ selection
                \end{tabular} 
            &   \begin{tabular}[c]{@{}l@{}}
                    Extension of the broken adaptive ridge \\method to recurrent events, involves \\repetition and reweighting of \\penalized $L_2$ models \\
                    Simultaneous variable selection and \\parameter estimation, accounts for \\clustering effects
                \end{tabular}       
            & \begin{tabular}[c]{@{}l@{}}Chronic septic \\granulomatosis \end{tabular} 
            &   \begin{tabular}[c]{@{}l@{}}
                    MSE \\
                    Number of predictor \\variables selected \\correctly, and number of \\predictor variables \\selected incorrectly
                \end{tabular} 
            & Yes \\
            \hline 
            \#3 & 2019
            & \begin{tabular}[c]{@{}c@{}}Wang, \\\textit{et al}.\end{tabular}
            &   \begin{tabular}[c]{@{}l@{}}
                    Machine Learning for \\Survival Analysis: \\A Survey 
                \end{tabular}
            &   \begin{tabular}[c]{@{}l@{}}Literature \\review\end{tabular} 
            &   \begin{tabular}[c]{@{}l@{}}
                    Introduction to survival analysis, \\overview of classical methods and \\of learning methods \\Recurrent events are mentioned, but \\ML methods are not developed
                \end{tabular} 
            & / & / & / \\
            \hline 
            \#4 & 2019 
            & \begin{tabular}[c]{@{}c@{}}Gupta, \\\textit{et al}.\end{tabular}
            &   \begin{tabular}[c]{@{}l@{}} 
                    CRESA: A Deep \\Learning Approach to \\Competing Risks, \\Recurrent Event \\Survival Analysis
                \end{tabular} 
            &   \begin{tabular}[c]{@{}l@{}}Deep \\neural \\networks\end{tabular} 
            &   \begin{tabular}[c]{@{}l@{}}
                    LSTM neural networks with the \\introduction of the cumulative \\incidence curve to take into account \\competitive and/or recurrent events 
                \end{tabular}
            & \begin{tabular}[c]{@{}l@{}} MIMIC III \\Machine failure \\data\end{tabular} & \begin{tabular}[c]{@{}l@{}}Harrell’s C-index \\MAE\end{tabular} & No \\
            \hline 
            \#5 & 2019
            & \begin{tabular}[c]{@{}c@{}}Jing, \\\textit{et al}.\end{tabular}
            &   \begin{tabular}[c]{@{}l@{}}A deep survival \\analysis method based \\on ranking\end{tabular} 
            &   \begin{tabular}[c]{@{}l@{}}Deep \\neural \\networks\end{tabular} 
            &   \begin{tabular}[c]{@{}l@{}} 
                    Extension of the DeepSurv model \\(neural networks for competitive \\events) with the use of ranking in the \\loss function on the differences \\between observed and predicted \\values 
                \end{tabular} 
            & \begin{tabular}[c]{@{}l@{}}Myocardial infarction \\Breast cancer omic data\end{tabular} 
            & Harrell’s C-index & Yes \\
            \hline 
            \#6 & 2020
            & \begin{tabular}[c]{@{}c@{}}Bull, \\\textit{et al}.\end{tabular}
            &   \begin{tabular}[c]{@{}l@{}} 
                    Harnessing repeated \\measurements of \\predictor variables for \\clinical risk prediction: \\a review of \\existing methods
                \end{tabular} 
            & \begin{tabular}[c]{@{}l@{}}Literature \\review\end{tabular} 
            &   \begin{tabular}[c]{@{}l@{}}
                    Summary of existing methodology to \\provide clinical prediction depending \\on the nature on input data \\Both statistical and learning \\approaches are described, but no \\ML methods for recurrent events \\highlighted
                \end{tabular} 
            & / & / & / \\
            \hline 
            \#7 & 2021 
            & \begin{tabular}[c]{@{}c@{}}Kim, \\\textit{et al}.\end{tabular}
            &   \begin{tabular}[c]{@{}l@{}}
                    Deep Learning-Based \\Prediction Model for \\Breast Cancer \\Recurrence Using \\Adjuvant Breast Cancer \\Cohort in Tertiary \\Cancer Center Registry 
                \end{tabular} 
            &   \begin{tabular}[c]{@{}l@{}}Deep \\neural \\networks\end{tabular} 
            &   \begin{tabular}[c]{@{}l@{}}
                    Use of Weibull Time To Event \\Recurrent Neural Network, an \\extension of recurrent neural \\network to sequentially estimate \\time to next event
                \end{tabular} 
            & \begin{tabular}[c]{@{}l@{}}Breast cancer \\registry in \\Korea \end{tabular} 
            & \begin{tabular}[c]{@{}l@{}}Harrell’s C-index \\MAE\end{tabular} & Yes \\                                
          \bottomrule
      \end{tabular} 
      \\
      LSTM = long short-term memory; MAE = mean absolute error; ML = machine learning. Articles were sorted by publication year. \#1 to \#3 were identified via hand searches and \#4 to \#7 via Pubmed.
    \label{tab:table1}
    \end{table}
\end{landscape}

\newpage
%%%%%%%%%%%%%%%%%%%%%%%%%%%%%%%%%%%%%%%%%%%%%%%%%%%%%%%%%%%%%%%%%%%%%%%%%%%%%%%%%%%%%%%%
\section{Materials and method}
Standard statistical models were not designed to handle high dimension. The most commonly used ones were described and compared to approaches identified in the literature review. The latter were detailed in the coming sections. Only published approaches with available code were considered (Table 1). Measures of performance for evaluation and comparison were described and were selected in order to stick to survival framework, to specificities inherent to recurrent event data and to machine learning stakes. Finally, the simulation scheme was developed.

\subsection{Notations}
Let $\mathbf{X}_i$ be a $p$-dimensional vector of covariates, $\beta$ the associated regression coefficients, $\lambda_0(t)$ the baseline hazard function, $Y_i(t)$ an indicator of whether subject $i$ is at risk at time $t$, $\delta_i = 1$ when the subject experienced the event (else $0$). Let $E_i$ and $C_i$ be the time to event or censoring, $T_i = E_i \land C_i$ for the patient $i$, with $a \land b = min (a,b)$. $N_i^* (t$) denoted the number of events over the interval $[0, t]$. Of note, $i = 1, …, n$, with $n$ the number of subjects and $\mathbf{X} \in \mathbb{R}^{n*p}$ denoted the covariates matrix for all subjects.

\subsection{Modeling recurrent events}
\subsubsection{Standard statistical models}
Andersen-Gill (AG), Prentice, William et Peterson (PWP), Wei-Lin-Weissfeld (WLW) and the frailty models were developed as extensions of the Cox model \citep{andersen_coxs_1982, prentice_regression_1981, wei_regression_1989,vaupel_impact_1979, cox_regression_1972}. These methodologies commonly used models to handle recurrent event data. Their characteristics are summarized in Table \ref{tab:table2}. Further details on time scales and how models accounted for subject at risk can be found in Appendix.

\begin{table}[h]
	\caption{Standard Statistical Models for Recurrent Events Analyses}
    \begin{tabular}{ll}
    	\toprule
        & \\
        \textbf{Model} & \textbf{Components and specificities}  \\ 
        & \\
        \hline \hline          & Conditional model, accounts for the counting process as a time scale and unrestricted set for subjects at risk \\ 
        AG               & Recurrent events within individuals are independent and share a common baseline hazard function \\ 
                         & Intensity of the model: $\lambda_i(t)=Y_i(t)*\lambda_0(t)*\exp(⁡\beta^t X_i )$ \\
        \hline           & Conditional model, counting process as time scale and restricted set for subjects at risk \\ 
        PWP              & Stratified AG, stratum $k$ collects all the $k$th events of the individuals \\ 
                         & Hazard function: $\lambda_{ik}(t)=Y_i(t)*\lambda_{0k}(t)*\exp(⁡\beta_k^t X_i )$ \\
        \hline           & Marginal model, also stratified, calendar time scale and semi-restricted set for subjects at risk \\
        WLW              & Intra-subject dependence \\
                         & Hazard function: $\lambda_{ik}(t)=Y_i(t)*\lambda_{0k}(t)*\exp(⁡\beta_k^t X_i )$ \\
        \hline           & Extension of AG model \\
        Frailty          & Random term $z_i$ for each individual to account for unobservable or unmeasured characteristics \\
                         & Intensity of the model: $\lambda_i(t)=Y_i(t)*\lambda_0(t)*z_i*\exp(⁡\beta^t X_i )$ \\
        \bottomrule
    \end{tabular}
	\label{tab:table2} \\
	AG = Andersen-Gill; PWP = Prentice, William et Peterson; WLW = Wei-Lin-Weissfeld.
\end{table}

\subsubsection{Learning algorithms for variable selection}
A common approach to address high-dimension challenge is variable selection. Penalizing models help to reduce the space of parameter coefficients, which is called shrinkage. Widely used for regression and classification problems, Lasso penalization accepts null coefficients to select variables and Ridge helps to deal with multicollinearity in the data. Both penalization approaches have been extended to Cox models in standard survival analysis framework \citep{tibshirani_lasso_1997, simon_regularization_2011}. The purpose is to solve a constrained optimization problem of the partial log-likelihood of the Cox model, which is written

\begin{equation}
    \mathcal{L}(\beta)=\sum_{i=1}^n \delta_i \beta^t X_i - \sum_{i=1}^n \delta_i * \log\sum_{j\in \mathcal{R}(\tau_i)}^n \exp(\beta^t X_i)
\end{equation}

With $\mathcal{R}(t)$ the set of individuals who are “at risk” for the event at time $t$. For CoxLasso, regularization is performed using an $L_1$ norm penalty and $\hat\beta = \argmax_\beta \mathcal{L}(\beta), \|\beta\|_1 \leq s$ and for CoxRidge an $L_2$ norm penalty and $\hat\beta = \argmax_\beta \mathcal{L}(\beta), \|\beta\|_2 \leq s$, with $s \geq 0$. The lower the value of $s$, the stronger the penalization. Hyperparameters, named penalty coefficients, are used to determine its value and enable to control the impact of the penalty. \\

\citet{zhao_variable_2018} proposed an extension of these methods to recurrent events by developing the broken adaptive ridge (BAR) regression. The first iteration consists of a penalized $L_2$ model

\begin{equation}
    \hat\beta^{(0)} = \argmin_\beta (-2*\mathcal{L}_{mod}(\beta) + \xi_n \sum_{j=1}^p \beta_j^2), \xi_n \geq 0 
\end{equation}

If penalization hyperparameter $\xi_n \geq 0$, this is a Ridge penalty, and if $\xi_n = 0$ then $\beta^{(0)}$ is not penalized. We update for each iteration $\omega$:

\begin{equation}
    \hat\beta^{(\omega)} = \argmin_\beta (-2*\mathcal{L}_{mod}(\beta) + \theta_n \sum_{j-1}^p \frac{\beta_j^2}{\hat\beta_j^{(\omega-1)}}), \omega \geq 1
\end{equation}

BAR estimates are defined by $\hat\beta = \lim_{k \to \infty} \hat\beta^{(\omega)}$. The estimator benefits from the oracle properties of both penalties for model covariate selection and estimation. Cross-validation is recommended to optimize values of hyperparameters $\xi_n$ and $\theta_n$. According to \citet{kawaguchi_surrogate_2020}, estimates are not sensitive to variations of $\xi_n$ and optimization can be performed only on $\theta_n$. In the absence of a consensual single measure on cross-validation under recurrent events, two values for $\theta_n$ were studied in this paper, thereby considering two separate models. 

\subsubsection{Deep neural network}
\begin{figure}[h]
	\centering
    \includegraphics[width=0.3\textwidth]{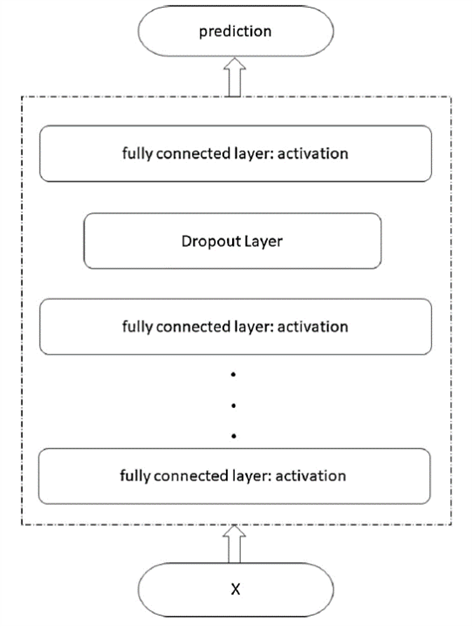}
    \caption{RankDeepSurv Neural Network Diagram (sourced from Jing 2019)}
	\label{fig:fig3}
\end{figure}

RankDeepSurv is a deep neural network proposed by \citet{jing_deep_2019} with fully connected layers (all neurons in one layer are connected to all neurons in another layer) (Figure \ref{fig:fig3}). 

The specificity of the RankDeepSurv neural network lies in the loss function adapted to survival, which results from the sum of two terms: one to constrain the survival model using an extension of the mean square error and the other to evaluate the rank error between observed and predicted values for two individuals. The loss function is written as

\begin{equation}
    L_{loss} (\theta)=\alpha_1*L_1 (\theta) + \alpha_2*L_2 (\theta)+ \mu * \|\theta \|_2^2 
\end{equation}

with
\begin{itemize}
    \item $\alpha_1, \alpha_2 > 0$ constant values, $\theta$ the weights of the network, $\mu$  regularization parameter for $L_2$  ;
    \item $L_1 = \frac{1}{n} \sum_{i=1,I(i)=1}^n (y_{j,pred}-y_{j,obs})^2, I(i)=1$ if $i$ is censored or if the predicted time to event is before observed time, else $0$;
    \item $L_2= \frac{1}{n} \sum_{I(i,j)=1}^n \big [ (y_{j,obs}-y_{i,obs}) - (y_{j,pred}-y_{i,pred}) \big ] ^2, I(i,j) = 1$ if $y_{j,obs}-y_{i,obs} > y_{j,pred}-y_{i,pred}$, else $0$.

\end{itemize}

Gradient descent is utilized for solving the minimization of $L_{loss}$. 

\subsection{Evaluation criteria to measure performance}
\subsubsection{Harrell's Concordance index}
Harrell’s C-index is a common evaluation criterion in survival analysis \citep{harrell_multivariable_1996}. This measure is the proportion of pairs of individuals for which the order of survival times is concordant with the order of the predicted risk. In the presence of censoring, the denominator is the number of pairs of individuals with an event. The C-index is estimated as follows

\begin{equation}
    \hat{\mathcal{C}} = \frac{\sum_{i \ne j} I(\eta_i < \eta_j) * I(T_i > T_j) * \delta_j}{\sum_{i \ne j} I(T_i > T_j) * \delta_j}
\end{equation}

With $\eta_i$ the risk of occurrence of the event. Of note, when two individuals are censored, then we cannot know which of the two has the event first. This pair is not included in the calculation. In the same way, if one of the individuals is censored and its censoring time is lower than the event time of another individual, we cannot know which of the two has the event first. This pair is also not included in the C-index calculation. If the C-index is equal to 1, it means a perfect prediction, and if the C-index $\leq 0.5$, it implies that the model behaves similarly or worse than random. Models with a higher C-index close to 1 are preferred. Harrell’s C-index was computed at each event.

\subsubsection{Kim's C-index}
\citet{kim_c-index_2018} proposed a measure of concordance between observed and predicted event counts over a time interval of shared observations. It is the proportion of pairs of individuals for whom the risk prediction and the number of observed events are concordant:

\begin{equation}
    \hat{\mathcal{C}}_{rec} = \frac{\sum_{i=1}^n \sum_{j=1}^n I \big ( N_i^*(T_i \land T_j) > N_j^*(T_i \land T_j) \big ) * I(\beta^tX_i > \beta^tX_j}{\sum_{i=1}^n \sum_{j=1}^n I \big ( N_i^*(T_i \land T_j) > N_j^*(T_i \land T_j) \big )}
\end{equation}

This extension of the C-index implies:

\begin{itemize}
    \item Two individuals are comparable up to the minimum time of follow-up;
    \item A pair contributes to the denominator if the two event counts are not equal.
\end{itemize}

As per Harrell's C-index in Equation (5), a score close to 1 indicates a better performance of the model. As opposed to Harrell’s C-index, Kim’s C-index was computed once across all the events.

\subsubsection{Error rate for active variables}
When simulating the datasets, the active status of each variable is known. Methods report the significant variables with a p-value < 0.05 (except deep neural networks). Significant variables are considered as positive tests for their active status. Some active variables likely have a false negative test ($FN$), and some passive variables have a false positive test ($FP$). The error rate ($err$) is the proportion of misclassified variables after prediction:

\begin{equation}
    err = \frac{FP + FN}{p}
\end{equation}

\subsection{Simulation scheme}
The following assumptions were made:

\begin{itemize}
    \item Active variables were continuous, and all have the same (non-zero) effect;
    \item The variables do not vary over time;
    \item Individuals were at risk continuously until end of follow-up;
    \item Censoring is not informative.
\end{itemize}

The generation of the covariates matrix, $X \sim \mathbb{N}_m (\mu,\sigma(\rho))$. $\mu = (\mu_1 ... \mu_p) =(a ... a)$ and $\sigma(\rho))$ was the covariance matrix with an autoregressive correlation structure and $\rho \in (0,1)$. The coefficients $\beta = (\beta_1 ... \beta_p) = (b,…,b,0,…,0)$ were associated with the $p$ covariates. $m$ coefficients were equal to a constant $b \in \mathcal{R}$ (the value of the active coefficients) and $p-m$ coefficients were equal to zero. The sparse rate was described by  $\frac{m}{p}$. The baseline hazard function followed a Weibull distribution with scale $\alpha>0$ and shape $\gamma>0$, and $\lambda_0(t) = \alpha * \gamma * t^{(\gamma-1)}$. The cumulated baseline hazard function could be expressed as $\Lambda_0(t)= \int_0^t \lambda_0(s)ds=\alpha*t^\gamma$. Hence the cumulative hazard function could be expressed as $\Lambda(t)=\Lambda_0 (t)*\exp⁡(\beta^t X_i)$. Conditional baseline hazard function was then defined as $\Tilde{\Lambda}_t(u) :=  \Tilde{\Lambda}^i(u | T_{i-1}) = \Lambda(u+t) - \Lambda(t)$. A frailty term $z_i$ i.i.d. was incorporated to account for heterogeneity. \\
\\
To maintain censoring rates, censored individuals were randomly drawn (censoring is not informative), as per \citet{jahn-eimermacher_simulating_2015}. The algorithm of \citet{jahn-eimermacher_comparison_2008} was applied to simulate event times $k$ for each subject $i$: 

\begin{equation}
    t_{i,1}=\Lambda^{-1} (t)(-\log⁡(\epsilon_1 )),  \\
    t_{i,k+1}=t_{i,1}+\Tilde{\Lambda}_{i,t_k}^{-1} (-\log(⁡\epsilon_{k+1}) )
\end{equation}

with $\epsilon_k \sim U[0,1 ]$. \\
\\
Train-test split was employed with a 70-30\% distribution. Datasets were generated with:

\begin{itemize}
    \item N = 100 subjects
    \item Censoring rate of 20\% 
    \item $\rho = 0.7$
    \item $b = 0.15$
    \ $\alpha = 1$ and $\gamma = 2$ 
    \item $z \sim Gamma(0.25)$ 
\end{itemize}

Scenarios include variations of the number of covariates p = 25, 50, 100, 150, and 200 and the sparse rate = 0\%, 25\%, and 50\%. For each of the 15 scenarios, 100 datasets were generated to account for variability.

%%%%%%%%%%%%%%%%%%%%%%%%%%%%%%%%%%%%%%%%%%%%%%%%%%%%%%%%%%%%%%%%%%%%%%%%%%%%%%%%%%%%%%%%
\section{Results}
\subsection{Data simulation overview}
\begin{figure}[H]
	\centering
    \includegraphics[width=0.65\textwidth]{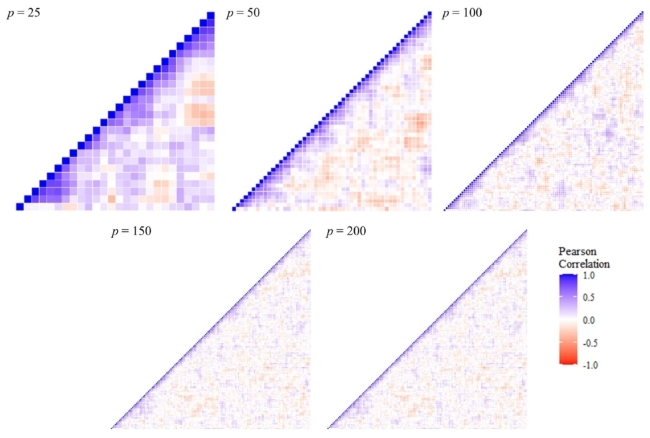}
    \caption{Heatmaps of Correlation with Variations of the Number of Variables (25, 50, 100, 150, 200)}
	\label{fig:fig4}
\end{figure}

\begin{figure}[H]
	\centering
    \includegraphics[width=0.6\textwidth]{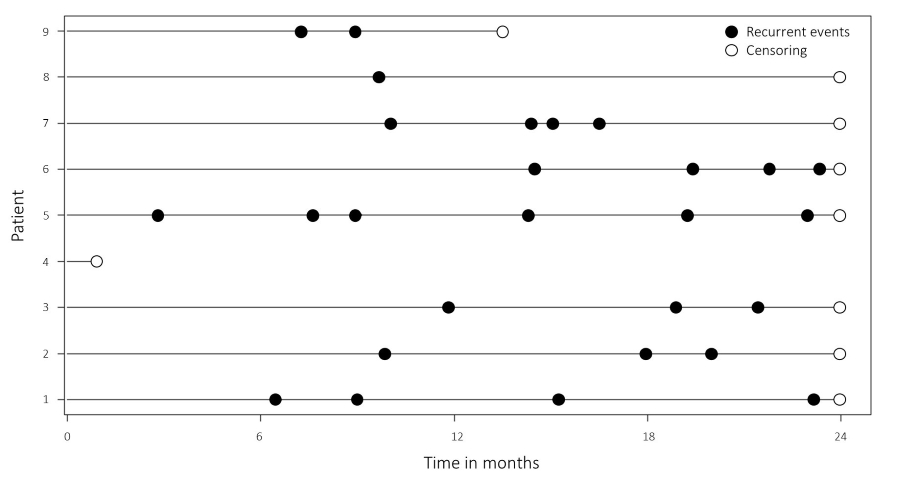}
    \caption{History of the  First Nine Simulated Individuals (for a given training set)}
	\label{fig:fig5}
\end{figure}

Datasets had identical characteristics in terms of number of individuals, structure of covariates, but differed across scenarios in terms of number of covariates and sparse rate. In the variance-covariance matrix, the covariates were highly correlated when they were close, then decreasingly close when they were further apart. Figure \ref{fig:fig4} captured this relationship across covariates with five datasets, regardless of the number of covariates. Figure \ref{fig:fig5} provided a visual representation of the history of nine individuals and their events over the follow-up period (and made it easier to understand Figure \ref{fig:fig1}).

\subsection{Evolution of average C-index}
\begin{figure}[H]
	\centering
    \includegraphics[width=1\textwidth]{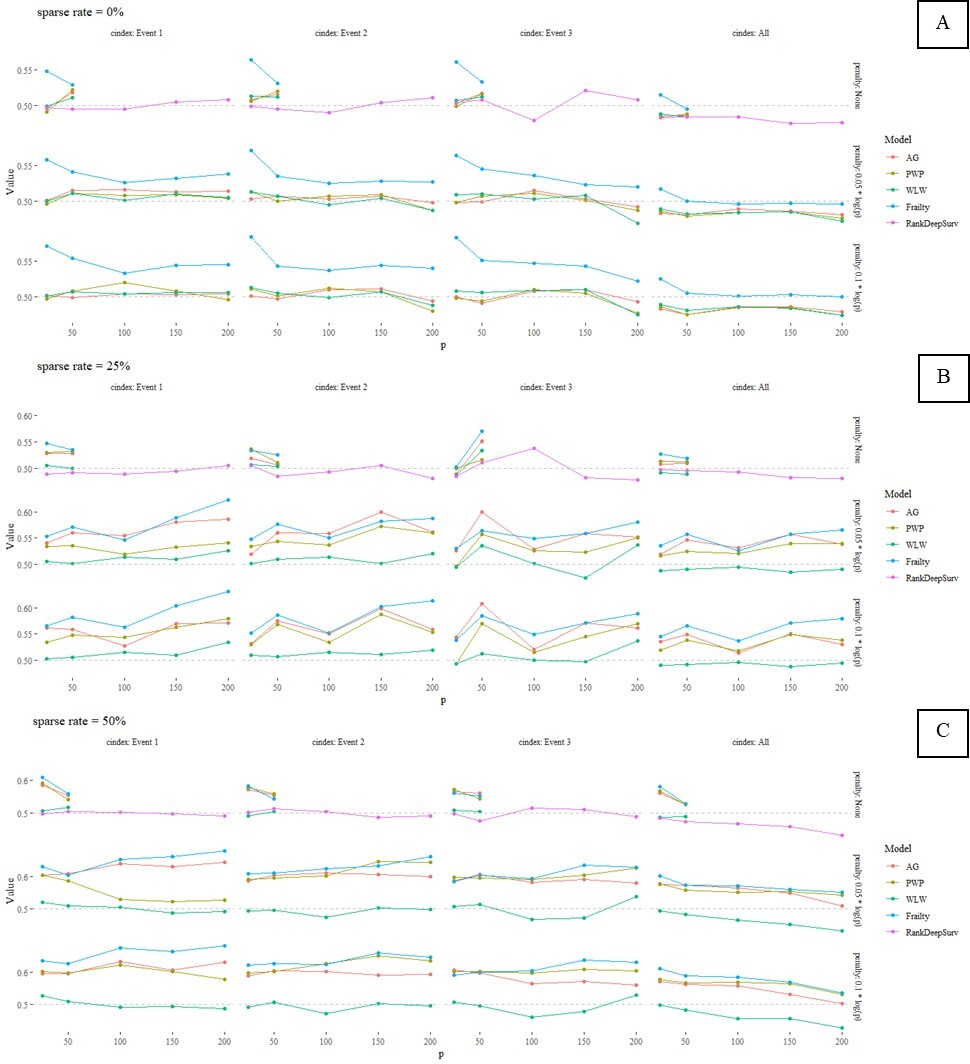}
    \caption{Evolution of Average C-indices with Sparse Rate Equal to 0\% (A), 25\% (B) and 50\% (C)}
	\label{fig:fig6}
\end{figure}

The evolution of average C-index was investigated across all 15 scenarios (Figure \ref{fig:fig6}). As expected, the standard models failed as soon as $p > n$. Whereas the C-indices were also expected to be around 0.5 when the sparse rate was zero, they increased as the sparse rate increased. The best performance was obtained using the frailty model. Other models showed similar trends, except for the WLW and RankDeepSurv models. The C-indices of these two models remained around the value of 0.5 (and even below) regardless of the scenario. The Kim’s C-index was more stable across the different number of covariates and sparse rates, although it tended to decrease as the number of covariates increased with sparse rate = 50\%. Small difference across penalty values was noticed as 0.05 penalized models and 0.1 penalized models followed similar trends.

\subsection{Focus on the variability of C-indices for two extreme scenarios}
Two extreme scenarios were thoroughly studied: one with no active variable and only 25 variables (A), and another on which models overall reported great performance with a sparse rate 25\% and over 150 variables (B). Similar trends in variability were observed across the two scenarios and for each C-index (Appendix Figure \ref{fig:fig8}). Kim’s C-index was less volatile across models and their penalties, with values ranging between 0.39 and 0.63 and 0.28 and 0.76 for (A) and (B), respectively. Harrell's C-index was increasingly variable in the first event (A: min = 0.26 and max = 0.74; B: min = 0.24 and max = 0.81), second event (A: min = 0.30 and max = 0.75; B: min = 0.17 and max = 0.85), and third event (A: min = 1 and max = 0).

\subsection{Error rate for active variables}
The results on the average error rates were displayed in Appendix Figure \ref{fig:fig9}. In the scenarios where the sparse rate was equal to 0\%, all models reported average error rates below 0.5, except penalized WLW with error rates around 0.75. Average error rates appeared similar when no penalty was applied. AG model had the lowest average error rate for each $p$, with a minimum value of 0.018 for the penalized model at $0.1*\log(p)$ and $p = 200$. The average error rate decreased when the penalty increased when $p > n$. For the scenarios with a sparse rate equal to 25\%, the unpenalized frailty model had the best performance, while the other models provided higher values. Similarly, penalties decreased the average error rate. Penalized AG models reported average error rates lower than 0.3. Finally, when the sparse rate was equal to 50\%, almost constant average error rates around 0.5 were observed for each model regardless of $p$.

%%%%%%%%%%%%%%%%%%%%%%%%%%%%%%%%%%%%%%%%%%%%%%%%%%%%%%%%%%%%%%%%%%%%%%%%%%%%%%%%%%%%%%%%
\section{Discussion}
The literature review provided in this paper enabled the identification of emerging approaches. A total of seven publications were found, testifying the shortage of available resources in this area. At the same time, these approaches had not been compared with one another. This may lead to erratic behavior and confusion when researchers wish to conduct robust and reliable analyses in such a context. This study thus proposed to evaluate some of the innovative learning algorithms which were developed to solve the high-dimensional framework when considering recurrent events. The investigation of the above 15 scenarios on simulated data arose specificities of both methodology and measures used for their evaluation of performance. \\
\\
Firstly, unpenalized standard approaches failed as soon as $p > n$ as expected, while penalized helped to improve their performance when $p < n$. This was typically expected as standard statiscal models were not designed for $p > n$ cases. AG and PWP models reported equivalent performance, while the frailty model consistently had the best performance. This was due to the construction of the frailty term from the simulation scheme. The WLW model performed mediocrely, regardless of if it was penalized or not; this was consistent with previous findings in the literature, exposing WLW models to be more appropriate with events of different types rather than recurrent events \citep{ozga_systematic_2018, ullah_statistical_2014}. Nevertheless, these models, each with their own specificities, can respond to differing needs, especially related to the research questions, but only methodological issues were meant to be illustrated in this study \citep{rogers_analysing_2014,amorim_modelling_2015,charlesnelson_how_2019}. \\
\\
Secondly, variable selection with penalties did not significantly increase performance, and few variables were even selected when the sparse rate was zero. Since only two values for the hyperparameter were explored, it seemed quite unlikely these would maximize the model performance. The deep neural network reported poorer performance; one reason could be that the format of the data was not suitable for the code.  \\
\\
Then, average error rates increased with the sparse rate. When the number of active variables was higher, the models tended to select the wrong variables. It seemed as if the models had a hard time learning and selecting the true active variables with a high sparse rate, while they managed to report better C-indices. This was linked to the variance-covariance structure chosen for data simulation.\\
\\
With regards to evaluation metrics, Kim’s C-index has shown higher stability and robustness compared to other metrics and stood for a criterion evaluating the entire set of events. Harrell's C, on the other hand, was measured at each event, making it difficult to interpret in terms of global performance. \\
\\
Nevertheless, some limitations should be noted. The literature review presented several drawbacks. Publications whose objective was variable selection without explicit dimension reduction, such as \citet{tong_variable_2009} and \citet{chen_variable_2013} could not be captured because of the elaborated research strategy. Also, it was not always easy to assess how the outcome was really considered, especially for neural networks which make little mention of the expected structure to process the data. Furthermore, as mentioned above, the lack of hyperparameter optimization for variable selection made BAR approach inconclusive. Also, a cross-validation would have highlighted the robustness of the results \citep{heinze_variable_2018}. Furthermore, other evaluation measures have been used in the literature, e.g., the mean square error, the mean absolute error, the log-likelihood \citep{wu_lasso_2013, zhao_variable_2018, ullah_statistical_2014}. An additional way to investigate active variables would be to assess the importance of the variables by permutation \citep{fisher_all_nodate}. By choosing a performance measure beforehand, this consists in permuting k times the order of the covariates and calculating k times the performance of the model. Finally, the simulations scheme itself presented several drawbacks. Covariates were not time-dependent and shared the same effect on the outcome, which may seem implausible in real life and made the interpretation of the results hard to generalize. Also, although the simulation of the data maintained censoring rates, it was not based on a distribution of censoring time, while one should be able to genuinely control them \citep{wan_simulating_2017, penichoux_simulating_2015}.

%%%%%%%%%%%%%%%%%%%%%%%%%%%%%%%%%%%%%%%%%%%%%%%%%%%%%%%%%%%%%%%%%%%%%%%%%%%%%%%%%%%%%%%%
\section{Conclusion}
To the author’s knowledge, this study was the first to compare standard methods, variable selection algorithms, and a deep neural network in modeling recurrent events in a high-dimensional framework. \\
\\
Progress in medical care is leading to the use of embedded artificial intelligence (AI) technologies. One illustration of this is the booming market for AI medical devices. Such systems are typically designed to prevent the occurrence of events either at the hospital, elderly care home or outpatient setting, etc. Should these events be likely to occur repeatedly, and should all available data/knowledge be captured, then thorough, robust and appropriate analysis of recurrent events is crucial \citep{berisha_digital_2021}. \\
\\
Overall, this work raises many concerns for recurrent event data analysis in high-dimensional settings and highlights the current need for developing further approaches and to assess their performance in a relevant manner.

%%%%%%%%%%%%%%%%%%%%%%%%%%%%%%%%%%%%%%%%%%%%%%%%%%%%%%%%%%%%%%%%%%%%%%%%%%%%%%%%%%%%%%%%
\section{Appendices}
\subsection{Data components for modeling recurrent events when using standard statistical approaches}
\subsubsection{Set of individuals at risk}
Standard statistical models described do not encounter for individuals at risk in the same way. This induces prior data management for appropriate application.

\begin{itemize}
    \item The set of individuals at risk for the $k$th event comprised individuals who were at risk for the event. Different definition existed for the set of individuals at risk, mainly based on baseline hazard function;
    \item The unrestricted set, in which each subject could be at risk for any event regardless of the number of events presented, at all-time intervals;
    \item The restricted set contained only the time intervals for the $k$th event of subjects who had already presented $k-1$ events;
    \item The semi-restricted set contained for the $k$th event the subjects who had $k-1$ or fewer events.
\end{itemize}

\subsubsection{Timescales}
Timescales also embody key components to address at the data management stage. Three common timescales are:

\begin{itemize}
    \item Calendar time, in which the times denotes the time since randomization/beginning of the study until an event occurs;
    \item Gap time, or waiting scale, resets the time to zero when an event occurs, i.e., it corresponds to the time elapsed since the last event previously observed;
    \item Counting process is constructed as per calendar time, although it enables late inclusions and/or censoring.
\end{itemize}

\begin{figure}[h]
	\centering
    \includegraphics[width=0.65\textwidth]{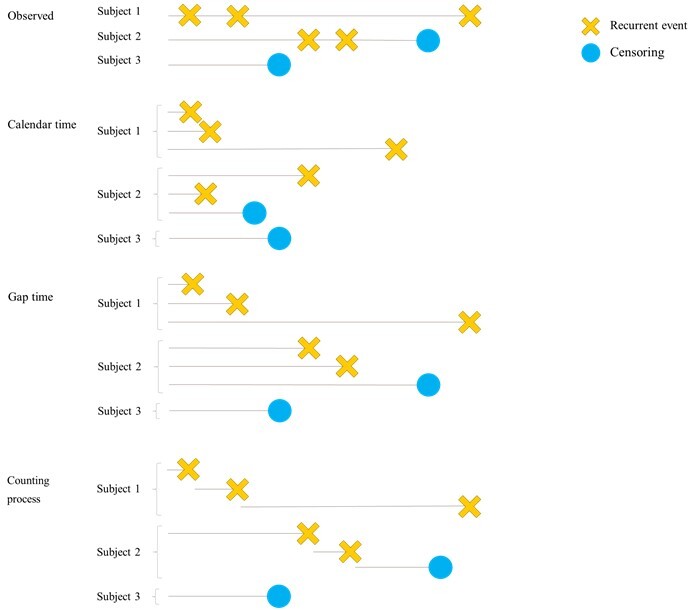}
    \caption{Timescales in Recurrent Events Analysis}
	\label{fig:fig7}
\end{figure}

\begin{figure}[H]
	\centering
    \includegraphics[width=1\textwidth]{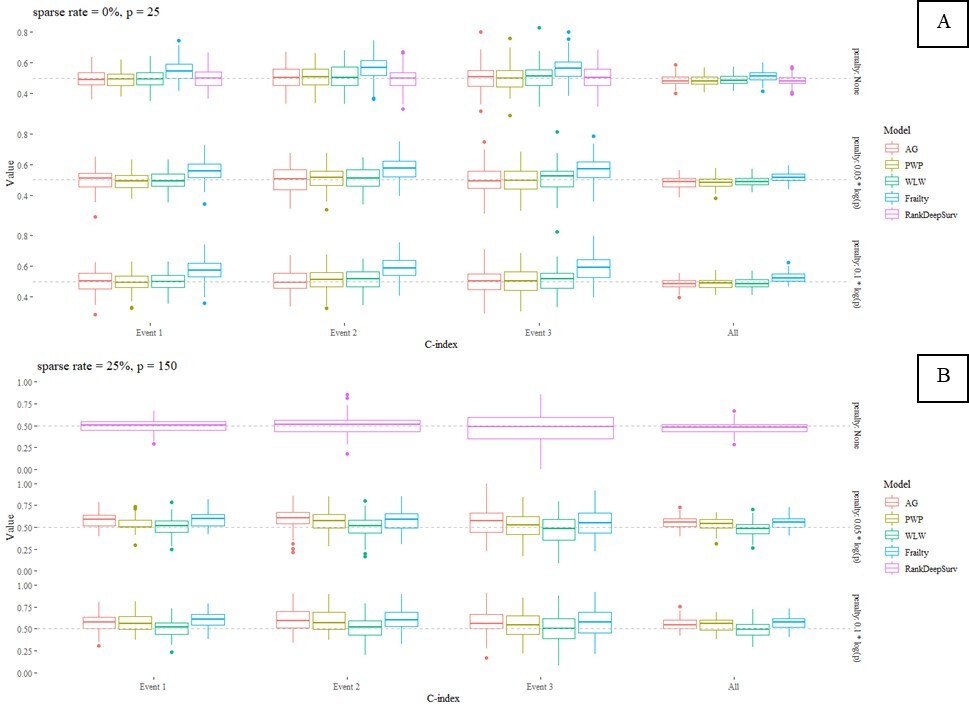}
    \caption{Variability of C-indices for Two Extreme Scenarios: Sparse Rate = 0\%, $p = 25$ (A) and Sparse Rate = 25\%, $p = 150$ (B)}
	\label{fig:fig8}
\end{figure}

Illustrations for timescales were provided in Figure \ref{fig:fig7}.

\newpage
\subsection{Further graphics from results}

\newpage
\begin{figure}[H]
	\centering
    \includegraphics[width=1\textwidth]{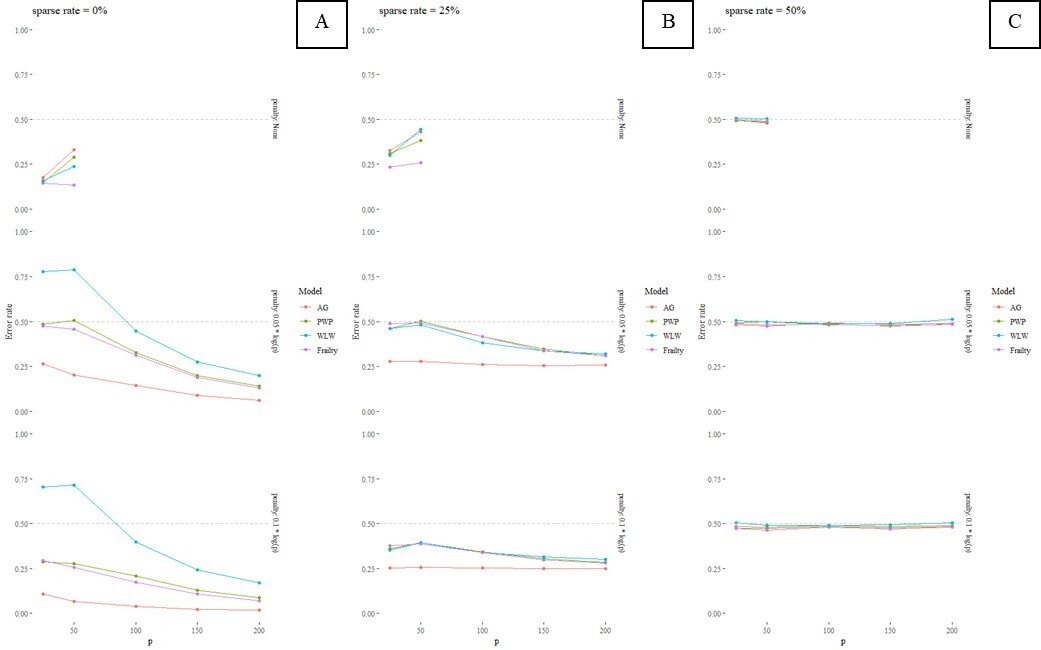}
    \caption{Evolution of Average Error Rates with Sparse Rate Equal to 0\% (A), 25\% (B) et 50\% (C)}
	\label{fig:fig9}
\end{figure}

\newpage
\bibliographystyle{plainnat}
\bibliography{biblio}  %%% Uncomment this line and comment out the ``thebibliography'' section below to use the external .bib file (using bibtex) .





% \end{thebibliography}

\end{document}